**Graphene oxide membranes: on the absence of "graphene capillaries", "ultrafast flow rate" and "precise sieving".***

Alexandr Talyzin

*Umeå University, Department of Physics, S-90187 Umeå, Sweden. e-mail: alexandr.talyzin@umu.se*

**Abstract.**

The data reported by Q.Yang et al[1] suggest absence of precise "sieving" and ultrafast flow of solvent across graphene oxide (GO) membranes. The "ultrafast flow" is not experimental fact but result of earlier speculated unrealistic geometrical model which assumed close packed arrangement of hole-free micrometer sized GO flakes with only 0.1% of area in each layer available for permeation. The data by Q.Yang et all[1] demonstrate that at least several percent of total layer area are available for permeation is real GO membranes due to holes between irregular shape of flakes. At least 2-3 percent of area also needs to be added to the permeation cross section due to holes and cracks in GO flakes, especially abundant when enforced by prolonged sonication of dispersions. Permeation of solutions mostly through pinholes which penetrate through tens of GO layers[1] suggests that previously proposed "graphene capillaries" are not required to explain water flow across the membrane. Taking into account more realistic packing of irregularly shaped GO flakes with holes between the flakes and across the flakes, 2-3 orders of magnitude shorter permeation path should be formed as compared to earlier suggested geometrical model with close packed square GO flakes. Considering shorter zigzag permeation path between hydrophilic GO sheets and 5-10% of area in each layer available for permeation due to holes across and between GO flakes, permeation rates across GO membranes can be explained by trivial diffusion. In absence of "graphene capillaries", "ultra-precise sieving" related to the "cutoff" value of 4.5Å (hydration diameter of ions) provided by "graphene capillaries" also has little meaning. As it is well known from earlier studies and now demonstrated by Q.Yang et al[1] using their own samples the swelling of GO membranes (which controls size of permeation channels) is different for different solutions, depends on concentration of solutes and results in delamination of membranes in many solutions. In fact, Q.Yang et al[1] provide controversial data for stability of their GO membranes even in pure water citing use of special surfactant to prevent dissolving of their membranes. Finally, oxidation state of real GO membranes studied by Q.Yang et al[1] and other studies from the same group (2012-2018) remains to be unknown due to incorrect procedures used to analyze XPS spectra.

The paper by Q.Yang et al reported difference in permeation properties of thin membranes prepared using GO dispersions sonicated for long ("Conventional" GO) and short periods of time ("Highly Laminated" GO).[1]
Permeability of GO membranes to water vapors and impermeability by gases at moisture free conditions was first reported in 1956[2] and in 1961 H.P.Boehm et al [2, 3] provided first sketch for diffusion of water molecules across the "graphitic oxide" membranes with zigzag pathway between hydrophilic graphene oxide sheets (Figure 1a). Note that the term "graphene" was not yet known at that time and introduced by H.P.Boehm only in 1990-s. H.P.Boehm et all provided also first information about sieving properties of GO membranes "impermeable for substances of lower molecular weight".[3] These publications were brought to attention of R.Nair et al group back in 2012 following their first report vapor permeation properties of GO membranes.[4] However, none of their followed papers[1, 5, 6, 7] cited old studies.[2, 3] Note that high permeation rates reported in 1960-s by H.P.Boehm et all[3] were not named in his papers as superfast (similar permeation rates were cited for clay membranes) and the sieving was not named as "ultraprecise". Limitations of sieving were clearly identified in early studies since e.g. large alkaloid ions were found to permeate through membranes. Concentration dependence of swelling properties of GO in e.g. NaOH was also cited.[3]

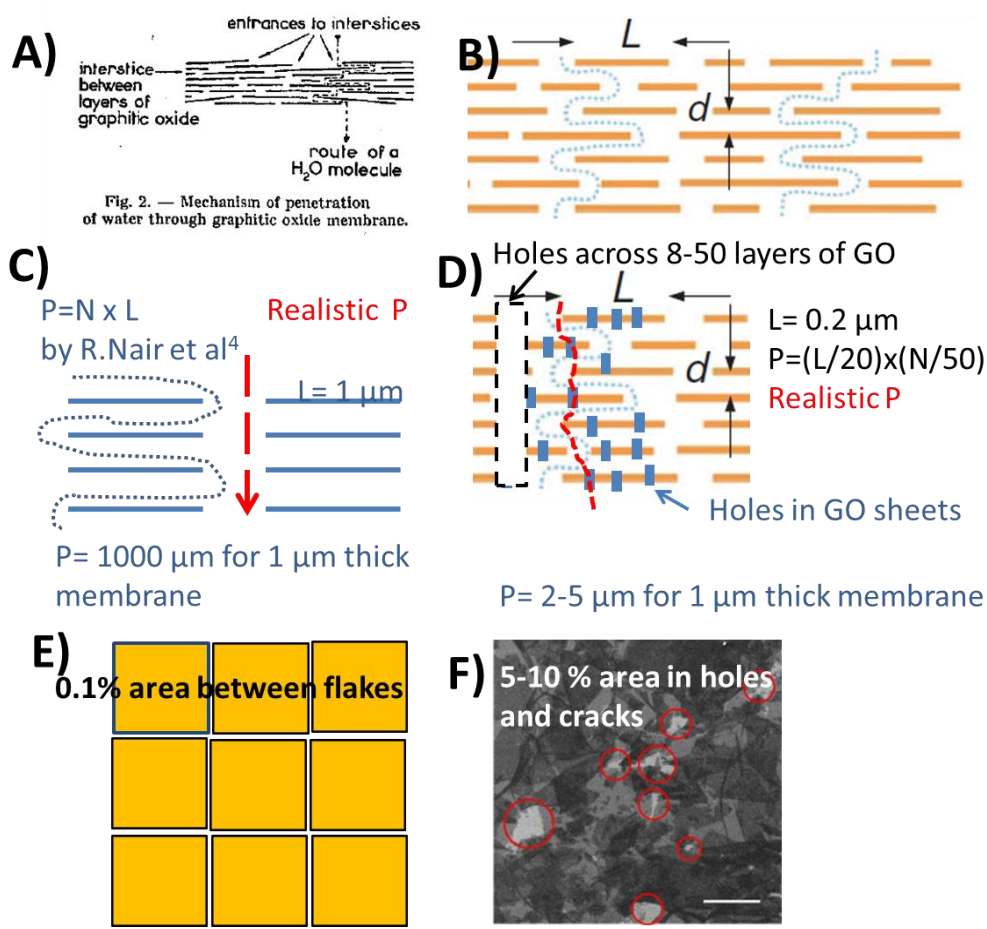

**Figure 1.** a) Permeation pathway proposed in 1961 by H.P.Boehm et al[3] b) Permeation path proposed by R.Nair et al in 2012[4] (reprinted with permission from AAAS) c) Scheme of the permeation path which corresponds to the calculation method proposed by R.Nair (P- pathway length, N- number of layers and L- size of flakes= 1µm), red line shows realistic alternative. d) Realistic pathway calculated using data provided in new study by Q.Yang et al[1] is about 500 times shorter. e) Geometrical model of GO flakes packing yielding permeation cross section of 0.1% of total area, size of flakes L=1 µm and distance between flakes 0.7 nm[4] f) Real packing of GO flakes in GO membranes showing micrometer sized holes and cracks (Supporting info Fig.9 in O.Yang et al,[1] )

The study by Nair et al reported similar water vapor permeation rate for GO membranes but speculated that the flow of solvent between GO sheets is "superfast".[4] In contrast to all previously existing knowledge about structure of graphene oxide, R.Nair et all suggested existence of large interconnected not oxidized areas on the surface of GO sheets and, citing the ref[4] "speculate that these empty spaces form a network of pristine-graphene capillaries within GO laminates".[4] It was speculated that graphene areas will not form graphitic stacking with ~3.3Å distance between layers but keep the same separation as between oxidized areas (~7-7.5Å). Moreover, it was speculated the increase in width of the graphene capillaries is identical to swelling of whole GO sheets thanks to "pillaring" effect by oxidized areas.[4] Later it was speculated that the swelling of GO is exactly the same for any solvents and solutions thus resulting in "ultraprecise sieving".[5] The "cutoff" related to the size of graphene capillaries in any solvents and solutions was speculated to be the same as for GO swelled in pure water (~4.5Å hydration radius).[1,5]

Theoretical modelling of permeation *across hydrophilic* GO membranes was then performed using diffusion *along* two *hydrophobic* graphene planes[4,5] and revealed e.g. that toluene and octanol molecules should not diffuse through water-filled 1 nm size capillaries.[5] In fact, both toluene and octanol are not miscible with water and will not diffuse through water in capillaries of any arbitrary size. Surprisingly, the models presented in ref[5] assumed full miscibility of water with toluene and octanol in bulk solutions.

It is remarkable, but none of the above cited speculations related to "graphene capillaries" were ever confirmed using characterization of the specific GO materials studied by this group in their experiments.[1, 4, 5, 6] Nevertheless, significant theoretical effort was involved in attempts to explain "superfast" flow of water in "graphene capillaries" using "frictionless flow", "large slip length" and "two-dimensional ice".[1, 4, 5, 6]

Here I argue that the "superfast flow" is result of incorrect geometrical model while there are no other arguments which would require existence of "graphene capillaries." The 2017 paper by Q.Yang et al[1] is the first attempt in 5 years (after first 2012 report[4]) where this group attempts to make detailed characterization of GO membranes (XPS, XRD in various solvents, flake size etc) but does not propose any analysis of new data in respect to their agreement/disagreement with earlier speculated models. In fact, the experimental data presented in the study by Q.Yang et al[1] provide decisive evidence for inconsistence of earlier proposed geometrical models with real structure of GO membranes, absence of "graphene capillaries" in GO and absence of "superfast flow" of solvent.

The velocity of water flow in the "channels" of GO membranes was calculated in R.Nair et al[4] using idealized geometrical model of membrane consisting of hole-free square-shaped 1 micrometer size GO sheets closely packed to each other with in-plane distance of only 0.7 nm between neighboring sheets. The model suggested that for each layer only 0.1% of area is available for flow of solvent and solvent must go around flakes on the way across the membrane (Figure 1 b).[4] The permeation path of water was estimated in this model to be about 1000 μm long for 1 μm thick membrane assuming 1 nm distance between the neighboring layers.

The permeation path was calculated in ref. 4 as N X L where N is number of layers and L is size of flake. However, this calculation suggests unrealistic permeation path shown in Figure 1 c, while the maximal length of water path between equally sized GO sheets is obviously L/2. Moreover, random stacking of flakes in the same model will result in average path of only ~L/4. That is 250 nm for 1 μm size flake. However, the average size of flakes determined for "conventional" type of GO membranes in new study by Q.Yang et al[1] is only about 200 nm (Supplementary Figure 1) thus making plausible path of L/4 to be about only 50 nm long. The total permeation path length in 1 μm thick membrane is then smaller by factor of 20 only taking into account real flake size of conventional GO membranes and realistic path around them. However, it is well known that GO flakes are perforated by numerous holes, especially if dispersions were strongly sonicated as in refs [4,5,7] It is known that additional holes and cracks are produced as a result of sonication.[8] Therefore, the distance between these holes is likely to be even smaller (~10-20 nm). As a result the total permeation path length in 1 μm thick membrane decreases from 1000 μm to ~10 μm. Moreover, the new study by Q.Yang et al [1] demonstrates that 8 nm-thick "highly laminated" membranes are permeated dominatingly through "pin-holes" between flakes (Figure 2b in ref [1]) and these flakes are not square shaped. The "pin-holes" were also found to influence permeation to at least 50 nm depth (Figure 3b in ref 1) even for membranes prepared without using strong sonication.

Therefore, it is very likely that permeation across "conventional" membranes prepared by standard sonication routines is dominated by the holes and cracks to much stronger extent, at least up to the thickness of 10-50 nm. This makes number of turns in the zigzag permeation pathway 10-50 times smaller than in the original geometrical model. [4] Assuming only 20 zigzag turns in 1 μm thick membrane instead of 1000 and 10-20 nm pathway between GO sheets in every interlayer results in a effective permeation of pathway of only 2-5 μm instead of originally speculated 1000 μm, (Figure 1d).

It is also easy to see that the area of holes between GO flakes is not 0.1% as in originally proposed system of close packed square shaped flakes[4] but accounts to at least 5-10% of total area for each layer of GO flakes (Figure 9 in supplementary file of ref [1]).

It can be concluded that the permeation path across the GO membranes was overestimated by 4-5 orders of magnitude in original models which resulted in speculation of "superfast flow" of water and motivated speculation of "graphene capillary".[4,5]

Assuming more realistic geometrical models and data collected by Q.Yang et al [1] from real materials the permeation rates across GO membranes can be explained *by trivial diffusion* between GO sheets in hydrophilic interlayers without "superfast" flow,[4] any need for "graphene capillaries", "frictionless flow with large slip length" and "square ice".[1, 4, 5]

Except for the "superfast flow" authors of refs[1,4,5,6,7] have not provided any other evidence or arguments which could support existence of "graphene capillaries" in GO membranes. In contrary, the absence of interconnected network of graphene capillaries in GO membranes is directly confronted by absence of He permeation in water free membranes[4] (size of "graphene channels" in water free membranes calculated as ~3Å in refs [1,4] allows permeation of gases), negligible BET surface area, direct HRTEM imaging,[9] conductivity mapping of GO sheets, analysis of Raman spectra,[10] invariance in permeation rates for laminates with different flake size[11] and many other methods which point out that small nm sized graphene areas are not interconnected until some reduction is applied.[12]

The existence of large interconnected graphene areas in GO was postulated in refs[1,4-8] but not supported by analysis of chemical composition of studied materials. New study by Q.Yang at al[1] speculates that 40-60% of GO surface "remains free from functionalization" but once again not supported by characterization data. The claim of interconnected "graphene" network of capillaries is supported by citation of review paper on GO reduction which clearly states that GO has "$sp^2$ clusters *isolated within* the $sp^3$ C–O matrix".[13] In fact, the ref 12 cited by Q.Yang et al[1] is review paper which mentions 60% of *carbon atoms* to be in not oxidized state (citing A.Lerf et al paper from 1998),[14] not providing any estimate for *areas* of oxidized/not oxidized regions.

The elemental composition of GO is of critical importance for verification of existence of large interconnected non-oxidized regions in GO. It is surprising, but the only XPS data which provided C/O ratio (2.8) of GO membranes studied by authors of ref [1] over past 5 years was published in our earlier report on swelling of GO membranes (this study included sample provided by R.Nair in 2012) in water-ethanol mixtures.[15] The study by Q.Yang et al[1] cites the same Hummers oxidation as in refs.[4, 15] for preparation of their samples but reports remarkably smaller degree of oxidation ( C/O=3.2-3.6). However, as it is discovered in communication with Prof. R.Nair related to first submission of this correspondence letter,[20] the C/O ratio reported in this study was estimated without recording oxygen part of XPS spectra using invalid procedure based on arbitrary deconvolution of only C1s part of spectra. As reported in the study the oxygen content of 23 ± 2%  was found for HLGO and 22 ± 2%. for CGO. This corresponds to 75.9%  and 79.2 % of  carbon taking into account provided C/O ratios. The sum of atomic concentration for carbon and oxygen are equal then to 98.9% and 101.2% respectively. No impurities are reported for both samples while for HLGO the sum of atomic concentrations exceeds 100%. Moreover, the interpretation of XPS spectra deviates from the commonly accepted.[16, 17] Three broad peaks observed in C1s spectra were fitted with four components assigned to C-C, C-O, C=O and CO(OH).[1] The C1s peak at about 287eV is typically assigned to C-O-C and C-OH functional groups in agreement with the structural model of Lerf-Klinowski.[14] This model suggests that C=O groups are found only on the edges of GO flakes. Note that even few percent's of C=O suggests existence of large amount of holes in GO planes. The 22-25% of C=O bonds reported in ref[1] suggest that a structure with completely disrupted graphene backbone. This interpretation of  XPS spectra was modified in the more recent publication by the same group[7] (following my communication with authors) emphasizing arbitrary character of the fitting which provides different C/O ratio for the same spectra. It can be concluded that C/O ratio and degree of oxidation for GO membranes studied by authors of ref's[1,4-7] in 2012-2018 remains to be unknown. Note that analysis of precursor graphite oxides used for preparation of membranes is not presented in new study and was never presented in previous publications from this group.[4-7]

The study by Q.Yang et al  reports also ultrafast and ultraprecise "sieving" with "cut off"  provided by the size of "graphene capillaries", this time for  20-30 nm thick GO membranes.[1] The "precise cut-off" for molecules and ions with hydration radius of 4.5Å was proposed in ref[5] assuming the size of "graphene capillaries" provided by swelling in pure water to be valid in any kind of solvent and solutions.  However, many studies demonstrated absence of precise correlation between size of ions/ molecules and permeation of GO membranes over past 50 years.[3] It is known that GO swells in broad range of polar solvents with lattice expansion up to tens of Å, dissolves  in basic solutions and that the swelling which controls size of permeation "channels" depends on the nature and concentration of solutions.[18, 19, 19] Thus the number of exceptions from the "ultraprecise" ion/molecule size cutoff of diffusion across the membrane (named as sieving in ref[1]) exceeds the number of solutes which might follow the rule.

 The paper by Q.Yang et al [1]  for the first time admits well known fact[14,18] that GO swelling in polar solvents is different from swelling in pure water and shows XRD patterns for 70 nm thick membrane(Figure 3 in ref 1).[1]  The exact d-spacings  for the swelled structures are not reported (~8-15Å range reading from the figure) as well as reference swelling properties for thicker membranes.  Moreover, the swelling is confirmed for solvents which were reported in earlier papers not to permeate through GO membranes (e.g. ethanol and acetone in ref. [4]). However, observation of broad range of interlayer distances in GO membranes due to swelling in different solvents (well known for many years, see e.g. our earlier studies[14,18]) do not lead authors of ref [1] to conclusion that the "cut-off"  value in sieving experiments needs to be established individually for each solvent, each mixture of solvents and verified for whole range of used concentrations.

Moreover, the values of d(001) in several solvents shown in Figure 3 as swelling in liquid solvents are not reliably established. According to details of experimental procedures described in supporting information, the membranes were taken out of solution in dry glovebox (which results in almost complete evaporation for most solvents within 2-3 minutes) and mounted inside of XRD cell "filled with some organic solvent vapor". Therefore, the dried (at least partly) membranes were exposed to solvent vapor for unknown period of time. The quickly recorded XRD patterns (~3 min) were possibly recorded at random vapor pressures far from saturation. If that is the case, these data should not be used for evaluation of "sieving" experiments performed in solvent immersed state.

Finally, the study by Q.Yang et al [1] provides contradictory statements about stability of their membranes in water. It is reported that "all the membranes were found stable in all tested solvents and water" and at the same time "the HLGO membrane breaks once the water comes into contact with the membrane. Therefore, we used a small amount of surfactant (0.6mg/ml sodium dodecyl benzene sulfonate". It is unclear from the text if the surfactant was used in all or only in some part of published "sieving" experiments while it might affect swelling of membranes or interact with feed solutions.

*The letter presented here is revised version of correspondence submitted to Nature Materials rejected on 5 of December 2018 and rejected without reviewing. Original submission made on 13 June 2018 was reviewed by one referee who stated that "it is very hard to imagine interconnected nanocapillaries consisting of graphene-like regions through whole layered GO membranes." Nevertheless the letter was rejected on 11$^{th}$ Nov 2018. Original version of letter, reply by R.Nair and reviewer report will be published elsewhere or available upon "reasonable request".*